\documentclass[12pt]{iopart}

\usepackage{iopams} 

\usepackage{hyperref}
\usepackage{latexsym}
\usepackage{graphicx, amssymb, float,color}

\usepackage{appendix}

\usepackage{cite}

\newcommand{\cbr}{C_{\mathrm{bro}}}
\newcommand{\calv}{C_{\mathrm{A}}}
\newcommand{\cart}{C_{\mathrm{a}}}
\newcommand{\cven}{C_{\bar{\mathrm{v}}}}
\newcommand{\cliv}{C_{\mathrm{liv}}}
\newcommand{\ctis}{C_{\mathrm{tis}}}
\newcommand{\cinh}{C_{\mathrm{I}}}
\newcommand{\cmeas}{C_{\mathrm{measured}}}
\newcommand{\cbag}{C_{\mathrm{bag}}}

\newcommand{\cw}{C_{\mathrm{water}}}

\newcommand{\poxy}{P_{\mathrm{O}_2}}
\newcommand{\pcd}{P_{\mathrm{CO}_2}}

\newcommand{\lliv}{\lambda_{\mathrm{b:liv}}}
\newcommand{\ltis}{\lambda_{\mathrm{b:tis}}}
\newcommand{\hen}{\lambda_{\mathrm{b:air}}}
\newcommand{\lmuca}{\lambda_{\mathrm{muc:air}}}

\newcommand{\qbr}{q_{\mathrm{bro}}}
\newcommand{\qliv}{q_{\mathrm{liv}}}
\newcommand{\qalv}{\dot{V}_{\mathrm{A}}}
\newcommand{\qc}{\dot{Q}_{\mathrm{c}}}

\newcommand{\pr}{k_\mathrm{pr}}
\newcommand{\met}{k_{\mathrm{met}}}
\newcommand{\dke}{k_{\mathrm{diff},1}}
\newcommand{\dkz}{k_{\mathrm{diff},2}}

\newcommand{\vbro}{\tilde{V}_{\mathrm{bro}}}
\newcommand{\valv}{\tilde{V}_{\mathrm{A}}}

\newcommand{\vtis}{\tilde{V}_{\mathrm{tis}}}
\newcommand{\vliv}{\tilde{V}_{\mathrm{liv}}}
\newcommand{\vbag}{\tilde{V}_{\mathrm{bag}}}

\newcommand{\di}{\mathrm{d}}

\newcommand{\lk}{\left(}
\newcommand{\rk}{\right)}
\newcommand{\abs}[1]{| #1 |}

\begin{document}

\title[A modeling-based evaluation of isothermal rebreathing]{A modeling-based evaluation of isothermal rebreathing for breath gas analyses of highly soluble volatile organic compounds}

\author{J~King$^{1,2}$,   
	K~Unterkofler$^{1,3}$,  
        	G~Teschl$^2$, 
        	S~Teschl$^4$,
        	P~Mochalski$^{1,5}$,
	H~Ko\c c$^{3}$,
        	H~Hinterhuber$^6$ and
        	A~Amann$^{1,7}$    }

\address{

$^1$ Breath Research Institute, Austrian Academy of Sciences,
Rathausplatz~4,
A-6850 Dornbirn, Austria

$^2$ Faculty of Mathematics, University of Vienna,
Nordbergstr.~15,
A-1090 Wien, Austria

$^3$Vorarlberg University of Applied Sciences,
Hochschulstr.~1,
A-6850 Dornbirn, Austria


$^4$ University of Applied Sciences Technikum Wien,
H\"ochst\"adtplatz~5,
A-1200 Wien, Austria

 $^5$ Institute of Nuclear Physics PAN, Radzikowskiego 152, PL-31342 
 Krak\'{o}w, Poland
 
 $^6$ Univ.-Clinic of Psychiatry, Innsbruck Medical University,
  Anichstr.~35,
 A-6020 Innsbruck, Austria

$^7$ Univ.-Clinic for Anesthesia, Innsbruck Medical University, 
Anichstr.~35,
 A-6020 Innsbruck, Austria
}

\ead{\mailto{anton.amann@i-med.ac.at}, \mailto{anton.amann@oeaw.ac.at}}

\begin{abstract} 
Isothermal rebreathing has been proposed as an experimental technique for estimating the \emph{alveolar} levels of hydrophilic volatile organic compounds (VOCs) in exhaled breath. Using the prototypic test compounds acetone and methanol we demonstrate that the end-tidal breath profiles of such substances during isothermal rebreathing show a characteristic increase that contradicts the conventional pulmonary inert gas elimination theory due to Farhi. On the other hand, these profiles can reliably be captured by virtue of a previously developed mathematical model for the general exhalation kinetics of highly soluble, blood-borne VOCs, which explicitly takes into account airway gas exchange as major determinant of the observable breath output.\\
This model allows for a mechanistic analysis of various rebreathing protocols suggested in the literature. In particular, it predicts that the end-exhaled levels of acetone and methanol measured during free tidal breathing will underestimate the underlying alveolar concentration by factor of up to~1.5. Moreover, it clarifies the discrepancies between \emph{in vitro} and \emph{in vivo} blood-breath ratios of hydrophilic VOCs and yields further quantitative insights into the physiological components of isothermal rebreathing and highly soluble gas exchange in general.
\end{abstract}


\submitto{Journal of Breath Research}

\maketitle

\section{Introduction}
\label{sect:intro}

Over the past decade, much advance has been made in the attempt to extract the diagnostic and metabolic information encapsulated in a number of endogenous volatile organic compounds (VOCs) appearing in human exhaled breath~\cite{amann2007,amannbook,miekisch2004}. However, the quantitative understanding of the relationships between breath concentrations of such trace gases and their underlying systemic levels clearly lags behind the enormous analytical progress in breath gas analysis. In particular, formal means for evaluating the predictive power of various breath sampling regimes are still lacking.

Recently, several efforts have been undertaken to complement VOC measurements with adequate physical models mapping substance-specific distribution mechanisms in the pulmonary tract as well as in the body tissues. Some major breath VOCs have already been investigated in this form, e.g., during rest and exercise conditions or exposure scenarios~\cite{kumagai2000,anderson2003,moerk2006,pleil2008,isoprene,King2010a,koc2011,ieee2011,beauchamp2011}. Such mechanistic descriptions of the observable exhalation kinetics give valuable insights into the relevance of the measured breath concentrations with respect to the endogenous situation and hence are mandatory to fully exploit the diagnostic potential of breath VOCs.

Within this context, the main focus of this article will be on a modeling-based review of isothermal rebreathing, which has been proposed as an experimental technique for estimating the \emph{alveolar} levels of hydrophilic exhaled trace gases~\cite{jones1983,ohlsson1990,ohara2008}. This class of compounds has been demonstrated to significantly interact with the water-like mucus membrane lining the conductive airways, an effect which has become known as \emph{wash-in/wash-out} behavior. For further details we refer to~\cite{anderson2003,anderson2006}. As a phenomenological consequence, exhaled breath concentrations of such highly water soluble substances tend to be diminished on their way up from the deeper respiratory tract to the airway opening. The resulting discrepancies between the ``true'' alveolar and the measured breath concentration can be substantial (even if breath samples are drawn in a strictly standardized manner employing, e.g., CO$_2$- or flow-controlled sampling) and will depend on a variety of factors, including airway temperature profiles and airway perfusion as well as breathing patterns~\cite{tsu1991}. \\

In particular, the above-mentioned effect considerably departs from the classical Farhi description of pulmonary inert gas exchange~\cite{farhi1967}, on the basis of which the operational dogma has been established that end-tidal air will reflect the alveolar level and that arterial concentrations can be assessed by simply multiplying this value with the blood:gas partition coefficient $\hen$ at body temperature. This ``common knowledge'' has first been put into question in the field of breath alcohol testing, revealing \emph{observable} blood-breath concentration ratios of ethanol during tidal breathing that are unexpectedly high compared to the partition coefficient derived in vitro~\cite{jones1983}. Similarly, excretion data (i.e., the ratios between steady state partial pressures in expired air and mixed venous blood) of highly water soluble compounds (including the MIGET test gas acetone) have been shown to underestimate the values anticipated by treating the airways as an inert tube~\cite{zwart1986,anderson2010}.\\

In a previously published mathematical model for the breath gas dynamics of highly soluble blood-borne trace gases, airway gas exchange is taken into account by separating the lungs into a bronchial and alveolar compartment, interacting via a diffusion barrier mimicking pre- and post-alveolar uptake~\cite{King2010a}. This formulation has proven its ability to reliably capture the end-tidal breath profiles as well as the systemic dynamics of acetone in a variety of experimental situations and will be used here for illuminating the physiological processes underlying isothermal rebreathing tests as carried out in the literature. For comparative reasons, the illustration in the sequel will mainly be limited to acetone, with possible extrapolations to other highly soluble VOCs indicated where appropriate.

\section{Methods}\label{sect:methods}

\subsection{Experiments}\label{sect:experimental}
Extensive details regarding our experimental setup are given elsewhere~\cite{King2009,King2010GC,King2010a}. Here, we will only briefly discuss the parts relevant in the context of isothermal rebreathing.\\

The rebreathing system itself consists of a Tedlar bag (SKC, Dorset, UK) with a volume of $\vbag=3$~l, that can directly be connected to a spirometer face mask covering mouth and nose. From the latter, \emph{end-tidal} exhalation segments are drawn into a Proton Transfer Reaction Mass Spectrometer (PTR-MS; Ionicon Analytik GmbH, Innsbruck, Austria), which allows for VOC detection and quantification on a breath-by-breath resolution as described in~\cite{King2009}. The rebreathing bag is warmed to 37~$\pm$~1$^{\circ}\mathrm{C}$ by means of a specially designed outer heating bag (Infroheat Ltd., Wolverhampton, UK), cf.~\cite{ohara2008}. Heating is intended to assist the thermal equilibration between the alveolar tract and the upper airways and prevents condensation and subsequent losses of hydrophilic VOCs depositing onto water droplets forming on the surface wall of the bag. End-tidal acetone concentrations are determined by monitoring the protonated compound at $m/z=59$ (dwell time~200~ms). Additionally, we routinely measure the mass-to-charge ratios $m/z=21$ (isotopologue of the primary hydronium ions used for normalization; dwell time~500~ms), $m/z=37$ (first monohydrate cluster $\mathrm{H_3 O}^+(\mathrm{H_2 O})$ for estimating sample humidity; dwell time~2~ms), $m/z=69$ (protonated isoprene; dwell time~200~ms), $m/z=33$ (protonated methanol; dwell time~200~ms) as well as the parasitic precursor ions $\mathrm{NH}_4^+$ and $\mathrm{O}_2^+$ at $m/z=18$ and $m/z=32$, respectively, with dwell times of 10~ms each. In particular, following~\cite{ohara2008}, the pseudo concentrations associated with $m/z=32$ determined according to standard PTR-MS practice (see, e.g.,~\cite[Eq.~(1)]{schwarzfrag}) will be used as a surrogate for the end-tidal oxygen partial pressure $\poxy$ (relative to an assumed nominal level at rest of 100~mmHg). Similarly, calibrated pseudo concentrations corresponding to $m/z=37$ are considered as an indicator for absolute sample humidity $\cw$, see~\cite{King2010a} for further details. Partial pressures $\pcd$ of carbon dioxide are obtained via a separate sensor.~\Tref{table:measparams} summarizes the measured quantities used in this paper. In general, breath concentrations will always refer to end-tidal levels. 
 
\begin{table}[H]
\centering \footnotesize
\begin{tabular}{|lcr|}\hline
 {\large\strut}Variable&Symbol&Nominal value (units)\\ \hline \hline 
{\large\strut}Acetone concentration &$\cmeas$ & 1 ($\mu$g/l)~\cite{schwarzace}\\
{\large\strut}CO$_2$ partial pressure &$\pcd$ & 40 (mmHg)~\cite{lumbbook}\\
{\large\strut}Water content &$\cw$ & 4.7 (\%)~\cite{hanna1986a}\\
{\large\strut}O$_2$ partial pressure &$\poxy$ & 100 (mmHg)~\cite{lumbbook}\\
\hline
\end{tabular}
\caption{Summary of measured breath parameters together with some nominal values in end-exhaled air during tidal breathing at rest.}\label{table:measparams}
\end{table}

In total, five~normal healthy male volunteers (age range 26-55 years, median 35 years) were recruited and gave written consent to participate in a single cycle rebreathing protocol as described in~\sref{sect:single}. All phenomenological results were obtained in conformity with the Declaration of Helsinki and with approval by the Ethics Commission of Innsbruck Medical University.

\subsection{Physiological model}\label{sect:model}
\begin{figure}[H]
\centering
\begin{tabular}{c}
\includegraphics[scale=1.15]{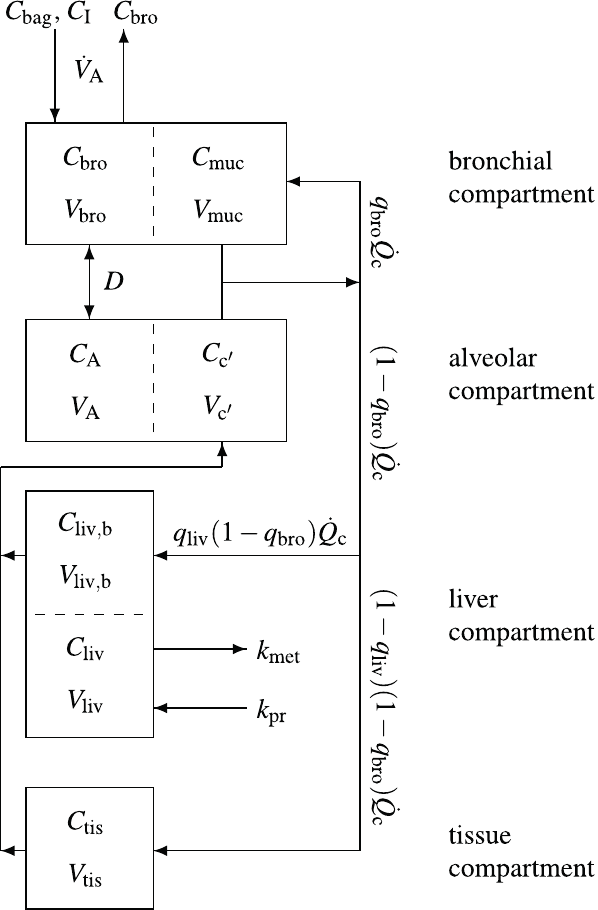}
\end{tabular}
\caption{Sketch of the model structure used for capturing dynamic VOC concentrations ($C$). Subscripts connote as follows: $\mathrm{bag}$--\emph{rebreathing bag}; $\mathrm{I}$--\emph{inhaled}; $\mathrm{bro}$--\emph{bronchial}; $\mathrm{muc}$--\emph{mucosal}; $\mathrm{A}$--\emph{alveolar}; $\mathrm{c'}$--\emph{end-capillary}; $\mathrm{liv}$--\emph{liver}; $\mathrm{tis}$--\emph{tissue}; $\mathrm{b}$--\emph{blood}. The conductance parameter $D$ reflects the net gas exchange between the bronchial and the alveolar tract over a complete respiratory cycle as described in the text. Note that $D$ is close to zero during free tidal breathing at rest (i.e., the measurable breath acetone levels are almost exclusively determined by the respective levels in the conducting airways) and varies in parallel with ventilatory flow, cf.~~\eref{eq:D}. In particular, higher values of $D$ correspond to an increased contribution of the alveoli to total pulmonary gas exchange.}\label{fig:model_struct}
\end{figure}

A schematic sketch of the model structure is presented in~\fref{fig:model_struct}. For the associated compartmental mass balance equations and the underlying model assumptions we refer to~\sref{sect:modeleq} as well as to the original publication~\cite{King2010a}. Briefly, the body is divided into four distinct functional units, for which the underlying concentration dynamics of the VOC under scrutiny will be taken into account: bronchial/mucosal compartment ($\cbr$; gas exchange), alveolar/end-capillary compartment ($\calv$; gas exchange), liver ($\cliv$; endogenous production and metabolism) and tissue ($\ctis$; storage). The nomenclature is detailed in~\tref{table:param}.

The measurement process is described by 
\begin{equation}\label{eq:meas}\cmeas=\cbr,\end{equation}
i.e., we assume that the measured (end-tidal) VOC concentration reflects the bronchial level. Dashed boundaries indicate a diffusion equilibrium, described by the appropriate partition coefficients $\lambda$, e.g., $\hen$. The nomenclature is detailed in~\tref{table:param}.

In particular, during tidal breathing at rest the present model replaces the familiar Farhi equation describing the steady state relationship between inhaled (ambient) concentration $\cinh$, measured breath concentrations $\cmeas^{\mathrm{F}}$, mixed venous concentrations $\cven$ and arterial concentrations $\cart$ during tidal breathing (cf.~\cite{farhi1967}),
\begin{equation}\label{eq:farhistst}
\cmeas^{\mathrm{F}}=\calv= \frac{\frac{\qalv}{\qc}\cinh +\cven}{\hen+\frac{\qalv}{\qc}}=\frac{\cart}{\hen},
\end{equation}
with the expression (cf.~\cite[Eq.~(21)]{King2010a})
\begin{equation}\label{eq:mystst}
\cmeas=\cbr=\frac{r_{\mathrm{bro}}\cinh+(1-\qbr)\cven}{(1-\qbr)z(\bar{T})\hen+r_{\mathrm{bro}}}=\frac{r_{\mathrm{bro}}\cinh+\cart}{z(\bar{T})\hen+r_{\mathrm{bro}}}.
\end{equation}
Here, $\qbr\approx 0.01$ is an estimate of the effective fractional bronchial perfusion (which might be substance-specific), while $r_{\mathrm{bro}}=\qalv/(\qbr\qc)$ denotes the associated \emph{bronchial} ventilation-perfusion ratio. Note that~\eref{eq:mystst} is only valid for highly water- and blood-soluble compounds such as acetone.\\

The term $z(\bar{T})\hen$ reflects an explicit temperature dependence of airway gas exchange that has been incorporated into the model. More specifically, the factor $z$ captures the change of VOC solubility in the airway mucosa and bronchial blood in response to fluctuations of the characteristic mean airway and bronchial blood temperature $\bar{T}$ (in~$\phantom{}^{\circ}$C). In~\cite[Eqs.~(3)-(5)]{King2010a} the latter has been shown to be estimable by virtue of the measured sample humidity $\cw$ and typically ranges around 34$\phantom{}^{\circ}$C during tidal breathing of room air at rest.
It is assumed that the temperature dependence of $\hen$ is proportional to the temperature change of the mucosa:air partition coefficient $\lmuca$ over the temperature range considered, i.e.~$z$ is given by 
\begin{equation}\label{eq:lmucb}
z(\bar{T}):=\lmuca(\bar{T})/\lmuca(37^{\circ}\mathrm{C}).
\end{equation}
Note that $z$ decreases with increasing temperature and takes values greater than one for $\bar{T}$ below body temperature. 

A particular convenience of the above representation stems from the fact that, as a first approximation, the mucosa layer can be assumed to inherit the physico-chemical properties of water. Correspondingly, $\lmuca$ can be estimated via the respective \emph{water:air} partition coefficient, which is usually available from the literature (see, e.g., the extensive compendium in~\cite{staudinger2001}).\\

A direct implication of the \emph{wash-in/wash-out} mechanisms discussed in the introduction is that during free tidal breathing at rest the amount of hydrophilic VOCs reaching the alveolar region with each inhalation will be approximately equal to the amount extracted with each exhalation. In other words, under these circumstances the bronchio-alveolar net exchange of such compounds over one breath cycle will be close to zero and a major part of the overall pulmonary gas exchange (e.g., up to 96\% as in the case of acetone~\cite{anderson2006}) can be ascribed to the conducting airways. As a consequence, a concentration gradient develops along the respiratory tree, with higher levels of the respective VOC found in the proximal (alveolar) parts. Increasing ventilatory flow rates and volume minimizes equilibration times with the more distal mucus layers and hence counteracts the saturation and depletion of the air stream during inhalation and exhalation, respectively. Consequently, the location of gas exchange is continuously shifted towards the alveolar region and the contribution of the latter becomes more pronounced. In particular, the end-exhaled (measured) breath concentration approaches values closer to the alveolar level.

Formally, this rationale can be incorporated into the model by including a diffusion term of the form $D(\cbr-\calv)$ into the underlying mass balance equations, cf.~\cite{King2010a}. This expression reflects the above-mentioned net gas exchange between the alveolar and the bronchial tract over a complete breath cycle. From the previous paragraph the conductance parameter $D$ will be close to zero during free tidal breathing at rest and increases with ventilatory flow, viz.,
\begin{equation}\label{eq:D}
D:=\dke \max\{0,V_T-V_T^{\mathrm{rest}}\}+\dkz \max\{0,\qalv-\qalv^{\mathrm{rest}}\},
\end{equation} 
where $\qalv$ and $V_T$ denote alveolar ventilation and tidal volume, respectively. The positive coefficients $k_{\mathrm{diff},i}$ can be estimated from experimental data, see~\cite{King2010a} and~\tref{table:param}.

A primary advantage of the above treatment is that much of the temporal and spatial heterogeneity associated with highly soluble gas exchange can be captured within a unidirectional rather than a bidirectional flow framework as in~\cite{anderson2003,kumagai2000}. From an operational view, this drastically reduces the number of required model equations, thereby yielding parsimonious parameterizations as well as tractable model structures.

\section{Results and discussion}\label{sect:res}

\subsection{Heuristic considerations}\label{sect:heuristics}

As has already been indicated in the introduction,~\eref{eq:farhistst} is inappropriate for capturing experimentally obtained arterial blood-breath concentration ratios (BBR) of highly water soluble trace gases during free breathing at rest (i.e., assuming $\cinh=0$). For instance, in the specific case of acetone, multiplying the proposed end-tidal population mean of approximately 1~$\mu$g/l~\cite{schwarzace} with a blood:gas partition coefficient of $\hen=340$~\cite{anderson2006} at body temperature appears to grossly underestimate arterial blood levels spreading around 1~mg/l~\cite{kalapos2003,wigaeus1981}. In contrast,~\eref{eq:mystst} asserts that the observable arterial blood-breath ratio in this case is 
\begin{equation}\label{eq:BBR}
\mathrm{BBR}=\frac{\cart}{\cmeas}=z(\bar{T})\hen+r_{\mathrm{bro}},
\end{equation} and will thus depend on airway temperature and airway blood flow as mentioned in the previous section. In particular, note that the BBR will be greater than the in vitro blood:gas partition coefficient $\hen$ for $\bar{T}$ below body temperature, cf.~\eref{eq:lmucb}. 

From~\eref{eq:BBR} it is clear that the more soluble a VOC under scrutiny, the more drastically its observable BBR will be affected by the current airway temperature, with an inverse relation between these two quantities. This deduction is consistent with measurements by~\cite{ohlsson1990} conducted in the field of alcohol breath tests, reporting a monotonous decrease of ethanol BBR values with increasing exhaled breath temperature. 
In the same contribution, BBRs of ethanol during normal tidal breathing were shown to typically exceed a value of 2500, which differs from the expected value $\hen=1756$ by more than 40\%. For comparison, based on a mean characteristic airway temperature of about $\bar{T}=34^{\circ}\mathrm{C}$~\cite{jones1982} and the water:air partition coefficient $\lmuca$ provided by~\cite[p.~568]{staudinger2001}, by substituting the parameter values in~\tref{table:param} we find that~\eref{eq:BBR} predicts an experimentally observable blood-breath ratio of ethanol in the range of 2250.~\Eref{eq:BBR} also suggests that in the case of VOCs with an extremely high affinity for both blood and water (e.g., ethanol, methanol), the discrepancies between in vivo and in vitro BBR values might largely be removed by warming the airways to body temperature before the breath sampling procedure. Contrarily, blood-breath ratios of less soluble VOCs, e.g., acetone, will additionally be affected by the comparatively large value of $r_{\mathrm{bro}}$.

Apart from providing some experimental evidence for the validity of~\eref{eq:BBR}, these ad hoc calculations suggest that the common practice of multiplying the measured breath concentration $\cmeas$ with $\hen$ to obtain arterial concentrations for highly soluble trace gases will result in an estimation that might drastically differ from the true blood level.

Isothermal rebreathing has been put forward as a valuable method for removing the above-mentioned discrepancies. 
The heuristic intention leading to isothermal rebreathing is to create an experimental situation where the breath levels of highly soluble VOCs are not altered during exhalation due to loss of such substances to the cooler mucus layer of the airways. This can be accomplished by ``closing the respiratory loop'', i.e., by continuous re-inspiration and -expiration of a fixed mass of air\footnote{For the sake of completeness, it should be noted that a small part of the air within the rebreathing circuit outlined in~\sref{sect:experimental} is actually consumed by the PTR-MS device (roughly $40$~ml/min), which, however, can safely be neglected in the present setting.} from a rebreathing receptacle (e.g., a Tedlar bag), causing the airstream to equilibrate with the mucosa linings over the entire respiratory cycle~\cite{ohlsson1990,anderson2006,ohara2008}. Additionally, warming the rebreathing volume to body temperature will ensure a similar solubility of these VOCs in both regions, alveoli and airways.

Formally, a model capturing the experimental situation during isothermal rebreathing can simply be derived by augmenting the model equations in~\sref{sect:modeleq} with an additional compartment representing the rebreathing receptacle, i.e.,
\begin{equation}\label{eq:bag}
\frac{\di\cbag}{\di t}\vbag=\qalv(\cbr-\cbag)
\end{equation}
and by setting $\cinh=\cbag$ in~\eref{eq:bro}. In the sequel, we will furthermore assume that the (mixed) venous blood concentrations stay roughly constant throughout the rebreathing periods investigated in the present context. This can be seen as a necessary physiological requirement backed by the fact that both blood and most body tissues have an enormous storage capacity with respect to hydrophilic compounds, implying that the corresponding concentrations within these compartments will vary extremely slowly in response to a change in inhaled concentration levels. Consequently, we may conclude that if the conducting airways are warmed to body temperature during rebreathing (i.e., if $z(\bar{T})$ in~\eref{eq:lmucb} approaches unity as temperature increases) the compartmental concentrations in the pulmonary tract will tend to a new \emph{quasi} steady state\footnote{We stress the fact that this is \emph{not} a steady state in the strict mathematical sense, i.e., corresponding to a situation where all time-derivatives of the mass balance equations in~\sref{sect:modeleq} are exactly zero. Effectively, in a hypothetical situation where rebreathing could be continued for several hours, breath acetone concentrations would still vary in parallel with a slow increase of body tissue and mixed venous blood levels.} obeying
\begin{equation}\label{eq:rebreath}\cmeas^{\mathrm{rebr}}=\cbr^{\mathrm{rebr}}=\cbag^{\mathrm{rebr}}=\calv^{\mathrm{rebr}}=\frac{\cart^{\mathrm{rebr}}}{\hen}= \frac{\cven}{\hen}.\end{equation}
This is a simple consequence of substituting $\cinh$ with $\cbr$ in~\eref{eq:mystst} according to the quasi equilibrium relation associated with~\eref{eq:bag}. Similarly, the classical Farhi model yields
\begin{equation}\label{eq:rebreathf}\cmeas^{\mathrm{F,rebr}}=\cbag^{\mathrm{F,rebr}}=\calv^{\mathrm{F,rebr}}=\frac{\cart^{\mathrm{F,rebr}}}{\hen}= \frac{\cven}{\hen}.\end{equation}
It is important to realize that if such quasi steady state conditions hold they will depend solely on the blood:air partition coefficient $\hen$ at 37$^\circ$C, thus rendering isothermal rebreathing as an extremely stable technique for providing a reproducible coupling between VOC levels in breath and blood. Particularly, it theoretically avoids the additional measurement of ventilation- and perfusion-related variables that would otherwise affect this relationship, thereby significantly simplifying the required technical setup for breath sampling. However, as will be illustrated in the following, the major practical obstacle is to guarantee that a quasi steady state as in~\eref{eq:rebreath} is effectively attained. \\

For the purpose of comparing the qualitative implications of~\eref{eq:farhistst} and~\eref{eq:mystst} it is instructive to note that the corresponding ratios between measured rebreathing concentrations and end-exhaled concentrations during free tidal breathing at rest will follow an entirely different trend. To this end, note that while from~\eref{eq:rebreathf} and by setting $\cinh=0$ in~\eref{eq:farhistst} we find that
\begin{equation}\label{eq:trendrebfarhi}
\frac{\cmeas^{\mathrm{F,rebr}}}{\cmeas^\mathrm{F}} \to \frac{\cven/\hen}{\cven/(\hen + \qalv/\qc)} =\frac{\hen+\qalv/\qc}{\hen},
\end{equation}
an analogous computation using~\eref{eq:rebreath} and~\eref{eq:mystst} shows that
\begin{equation}\label{eq:trendrebmod}
\frac{\cmeas^{\mathrm{rebr}}}{\cmeas} \to \frac{(1-\qbr)z(\bar{T})\hen+r_{\mathrm{bro}}}{(1-\qbr)\hen},
\end{equation}
where $z(\bar{T}) \geq 1$, cf.~\eref{eq:lmucb}.
For highly soluble trace gases, this observation constitutes a simple test for assessing the adequacy of the Farhi formulation regarding its ability to describe the corresponding exhalation kinetics. Indeed, for sufficiently large $\hen$, the right-hand side of~\eref{eq:trendrebfarhi} will be close to one, while the right-hand side of~\eref{eq:trendrebmod} suggests that rebreathing will increase the associated end-tidal breath concentrations. In other words, for this class of compounds a markedly non-constant behavior during the initial isothermal rebreathing period indicates that~\eref{eq:farhistst} will fail to capture some fundamental characteristics of pulmonary excretion. Such tests are of particular importance in the context of endogenous MIGET methodology (Multiple Inert Gas Elimination Technique, based on endogenous rather than externally administered VOCs~\cite{anderson2010}), as they might be used for detecting deviations of the employed gases from the underlying Farhi description (see also~\cite{King2010GC}).

\subsection{Single cycle rebreathing}\label{sect:single}
In this section we will discuss some simulations and preliminary experiments conducted in order to study the predictive value of isothermal rebreathing within a realistic setting. For this purpose we will first mimic isothermal rebreathing as it has been carried out by various investigators~\cite{jones1983,ohlsson1990}.

More precisely, rebreathing was instituted after $t_{\mathrm{start}}=2$~min of free tidal breathing at rest by inhaling to total lung capacity and exhaling until the bag was filled, thereby providing an initial bag concentration which can be assumed to resemble the normal end-exhaled level, i.e.,
\begin{equation}\cbag(0)=\cbr(0).\end{equation}
Rebreathing was then continued until either the individual breathing limit was reached or the CO$_2$ partial pressure increased above $55$~mmHg. The volume of the rebreathing bag was $\vbag=3$~l according to~\sref{sect:experimental}. \\

\begin{figure}[H]
\centering
\begin{tabular}{c}
\hspace*{-0.7cm}
\includegraphics[height=12cm]{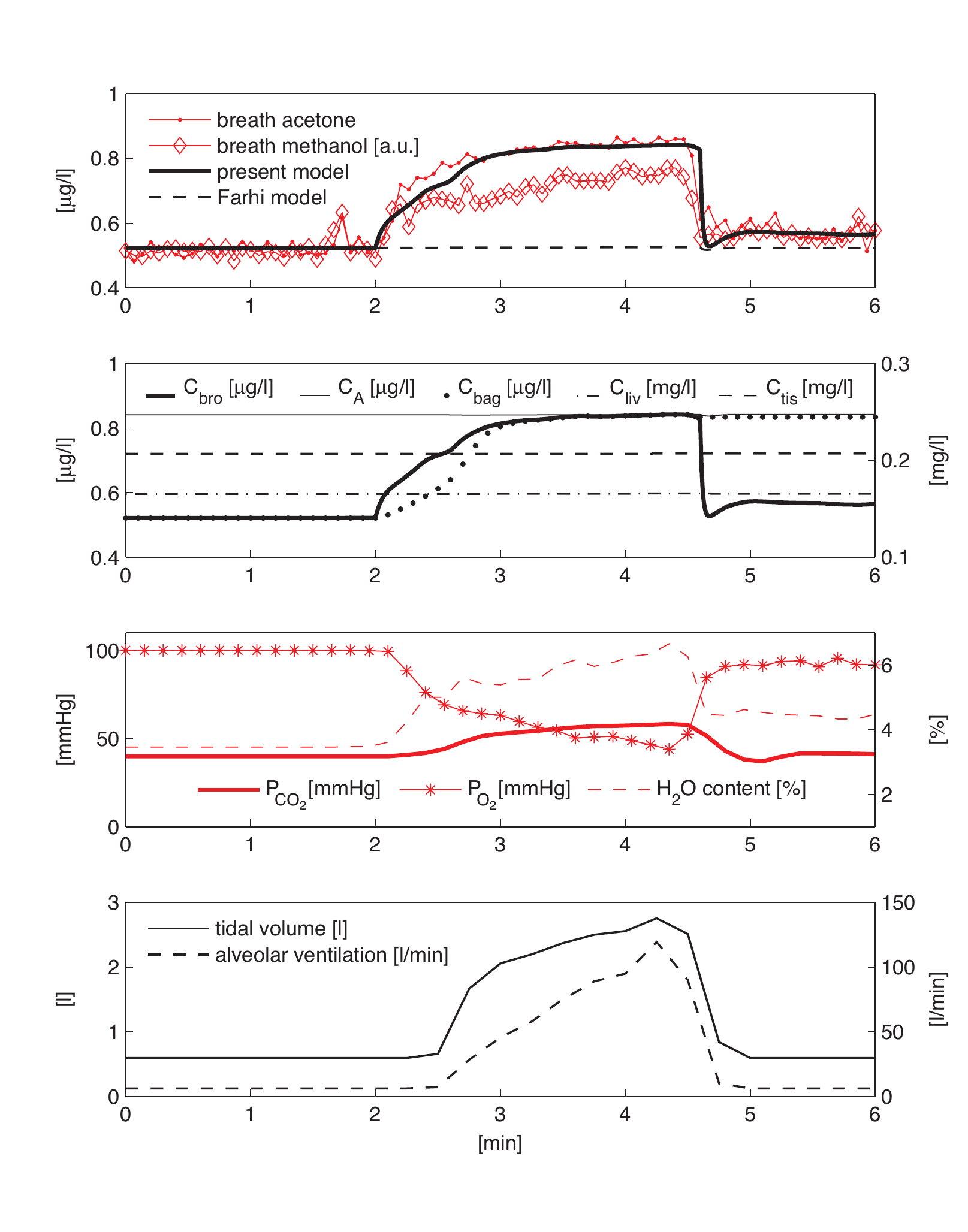}
\end{tabular}
\vspace*{-0.9cm}
\caption{Representative outcome of an isothermal rebreathing experiment during rest. Data correspond to one single normal healthy male volunteer from the study cohort in~\sref{sect:experimental} (Volunteer~1, see also~\fref{fig:rebreathing_single_all}). Isothermal rebreathing starts at $t_{\mathrm{start}}=2$~min and ends at approximately $t_{\mathrm{end}}=4.6$~min. Measured or derived quantities according to the experimental setup are shown in the first and third panel (red tracings), while data in the second and fourth panel correspond to \emph{simulated} variables as described in the text (black tracings).}\label{fig:rebreathing_single}
\end{figure}

\begin{figure}[H]
\centering
\begin{tabular}{c}
\hspace*{-0.7cm}
\includegraphics[height=14cm]{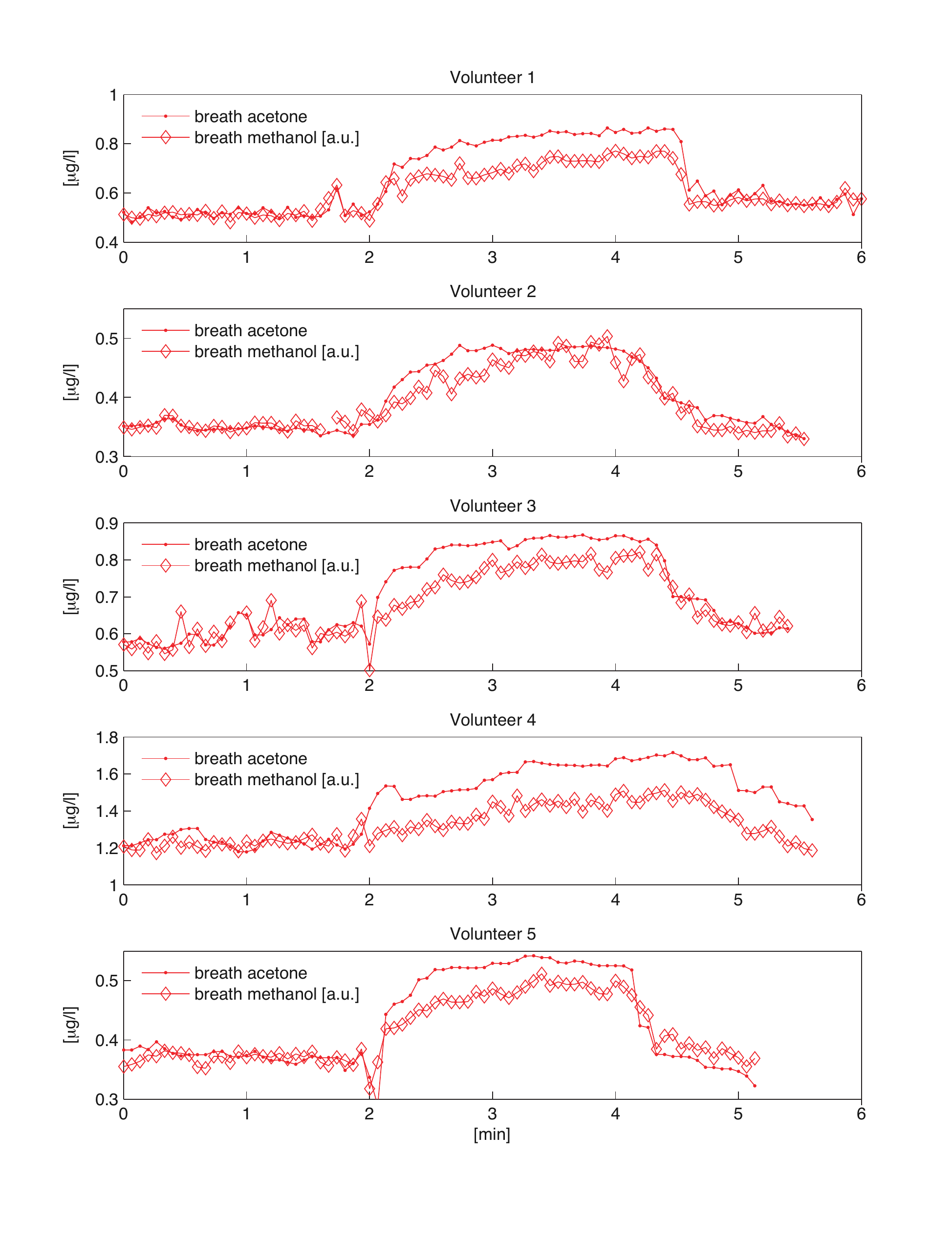}
\end{tabular}
\vspace*{-0.9cm}
\caption{Breath acetone/methanol profiles during isothermal rebreathing for all five volunteers in the study cohort. Breath methanol profiles were scaled to match the initial levels of breath acetone. Rebreathing starts at $t_{\mathrm{start}}=2$~min. As soon as the individual breathing limit is reached (roughly between 4 and 5~min), the volunteers disconnect the rebreathing bag and continue with free tidal breathing at rest.}\label{fig:rebreathing_single_all}
\end{figure}

Typical profiles of end-tidal acetone, water, CO$_2$ and oxygen content during normal breathing and isothermal rebreathing at rest are shown in~\fref{fig:rebreathing_single}. These representative data correspond to one single healthy male volunteer from the study cohort described in~\sref{sect:experimental}. An overview of the experimental outcome for all volunteers is given in~\fref{fig:rebreathing_single_all}. 

As has been explained above, $\poxy$ can be derived by scaling the end-tidal level of the PTR-MS pseudo concentration signal at $m/z=32$ to a basal value of 100~mmHg during free tidal breathing. Furthermore, breathing frequency $f$ and tidal volume $V_T$ may be simulated on the basis of the monitored values for $\pcd$ and $\poxy$. Under iso-oxic conditions, after a certain threshold value is exceeded, both frequency and tidal volume are known to increase linearly with alveolar $\pcd$. Moreover, the corresponding slopes (reflecting the chemoreflex sensitivity of breathing) are dependent on the current alveolar oxygen partial pressure~\cite{mohan1997,lumbbook}. More specifically, hypoxia during rebreathing enhances chemoreflex sensitivity, yielding a hyperbolic relation between the mentioned slopes and $\poxy$. These findings result in a simple model capturing the chemoreflex control of breathing in humans~\cite{duffin2000}, which has been re-implemented in order to compute $f$ and $V_T$ from basal values during free breathing, cf.~\fref{fig:rebreathing_single}. Here, for simplicity it is assumed that end-tidal $\pcd$ levels reflect those of the peripheral and central chemoreceptor environment. The ventilation rate follows from
\begin{equation}\label{eq:qalv}
\qalv=f(V_T-V_D), 
\end{equation}using an instrument (mask) deadspace volume of $V_D=0.1$~l. 

Since cardiac output can be expected to stay relatively constant during the rebreathing phase~\cite{ohara2008}, we fix its value at a nominal level of 6~l/min. Using the parameter values in~\tref{table:param} for a male of height 180~cm and weight 70~kg this completes the necessary data for simulating the above-mentioned rebreathing experiment. More specifically, the model response in the first panel of~\fref{fig:rebreathing_single} is obtained by integrating the differential equations~\eref{eq:bro}--\eref{eq:tis} and~\eref{eq:bag}, setting $\cinh =\cbag$ for $t \in [t_{\mathrm{start}},t_{\mathrm{end}}]$ and $\cinh \equiv 0$ otherwise. The initial concentrations for each compartment as well as the underlying endogenous acetone production rate $\pr =0.055$~mg/min for this specific experiment are derived from the algebraic steady state conditions at $t=0$.

From~\fref{fig:rebreathing_single}, the proposed model is found to faithfully reproduce the observed data. Further improvement of the goodness-of-fit might be achieved by employing parameter estimation techniques as described, e.g., in~\cite{isoprene,King2010a} which, however, would be beyond the scope of this paper.

In contrast, as can be anticipated from~\eref{eq:trendrebfarhi}, when using the classical Farhi formulation for gas exchange in the the pulmonary tract (i.e., when treating the conducting airways as an inert tube) the given breath acetone profile cannot be captured\footnote{Considering the fact that the Farhi formulation is included in the present model as a limiting case for $\qbr=0$ and $D \to \infty$~\cite[Fig.~4]{King2010a}, its associated output can be computed in a similar manner as described above.}. 

In particular, the presented data appear to consolidate the heuristic considerations in~\sref{sect:heuristics} and confirm that the alveolar concentration of acetone during free tidal breathing can differ from the associated bronchial (i.e., measured end-exhaled) level by a factor of more than~1.5. As has been explained in~\sref{sect:model}, this is due to an effective concentration gradient between the conducting airways and the alveolar space. During isothermal rebreathing, this gradient slowly vanishes and causes the measured breath concentration to approach the underlying alveolar concentration. It should be noted that the latter as well as the simulated levels in liver and storage tissue stay almost constant throughout this period, which is consistent with the quasi steady state assumption made in~\eref{eq:rebreath}. We also stress the fact that in order to simulate a similar breath profile by using the conventional Farhi formulation (i.e., by treating the airways as an inert tube), one essentially would have to postulate a temporarily increased endogenous acetone production during rebreathing, which, however, lacks physiological plausibility.

For comparative reasons, in~\fref{fig:rebreathing_single} and~\fref{fig:rebreathing_single_all} we also display the simultaneously obtained PTR-MS concentration profiles of breath methanol, scaled to match the initial levels of breath acetone. Taking into account a methanol blood:gas partition coefficient of $\hen=2590$ at body temperature~\cite{kumagai2000}, from~\eref{eq:trendrebmod} it can be deduced that for this compound the differences between concentrations extracted during free breathing and rebreathing primarily stem from the thermal equilibration between airways and alveolar tract. The corresponding rise in temperature is mirrored by a steady increase of sample water vapor $\cw$, approaching an alveolar level of about 6.2\%. 
In particular, the presented profiles for methanol show the necessity of including an explicit temperature dependence in models describing the rebreathing behavior of highly soluble VOCs.

As a last point in this section, we would like to emphasize the fact that all model parameters and initial conditions $p_i$ governing the simulation in~\fref{fig:rebreathing_single} can be ranked with respect to their impact on the observable breath acetone concentration during the isothermal rebreathing period $[t_{\mathrm{start}},t_{\mathrm{end}}]$ by numerically approximating the squared $L_2$-norm of the corresponding \emph{normalized sensitivities}, viz.,
\begin{equation}\label{eq:sens}
\varsigma(p_i):=\int\limits_{t_{\mathrm{start}}}^{t_{\mathrm{end}}}\lk\frac{\partial \cmeas(t)}{\partial p_i}\frac{p_i}{\max_s \abs{y(s)}}\rk^2 \mathrm{d} t.
\end{equation}
Adopting the nomenclature in~\sref{sect:modeleq}, these indices were calculated for all effective compartment volumes, partition coefficients, initial concentrations, and for $p_i \in \{\pr,\met,\qbr,\dke,\dkz\}$. Among these, the decisive quantities are the blood:gas partition coefficient (negative influence; $\varsigma(\hen)=2.07$) as well as the initial concentrations and partition coefficients for the liver and tissue compartment (positive influence; $\varsigma(\cliv(0))=0.06$; $\varsigma(\lliv)=0.06$; $\varsigma(\ctis(0))=1.39$; $\varsigma(\ltis)=1.39$). In particular, the latter essentially define the mixed venous blood acetone concentration $\cven$ according to~\eref{eq:cven}. As this concentration stays almost constant throughout the entire experiment, the above sensitivity analysis results are in direct agreement with~\eref{eq:rebreath}, predicting a rebreathing plateau that exclusively depends on $\hen$ and $\cven$. All other sensitivity indices take values below~0.015.

\subsection{Sequential rebreathing}\label{sect:rep}
On the basis of the results from the previous subsection, in the following we will briefly discuss a sequential rebreathing protocol developed by~\cite{ohara2008}. This regime aims at improving the patient compliance of conventional rebreathing by repeatedly providing cycles of five rebreaths (postulated to last approximately 0.5~min) with intermediate periods of free tidal breathing lasting approximately 10~min.
According to above-mentioned protocol, isothermal rebreathing is again instituted by inhaling to total lung capacity and exhaling to residual volume into a Tedlar bag with a volume of $\vbag=3$~l. After each rebreathing cycle, the bag is closed, a small amount of bag air ($< 100$~ml) is measured and the volunteer starts the next cycle by exhaling to residual volume and inhaling from the bag. \\

\begin{figure}[H]
\centering
\begin{tabular}{c}
\includegraphics[height=11cm]{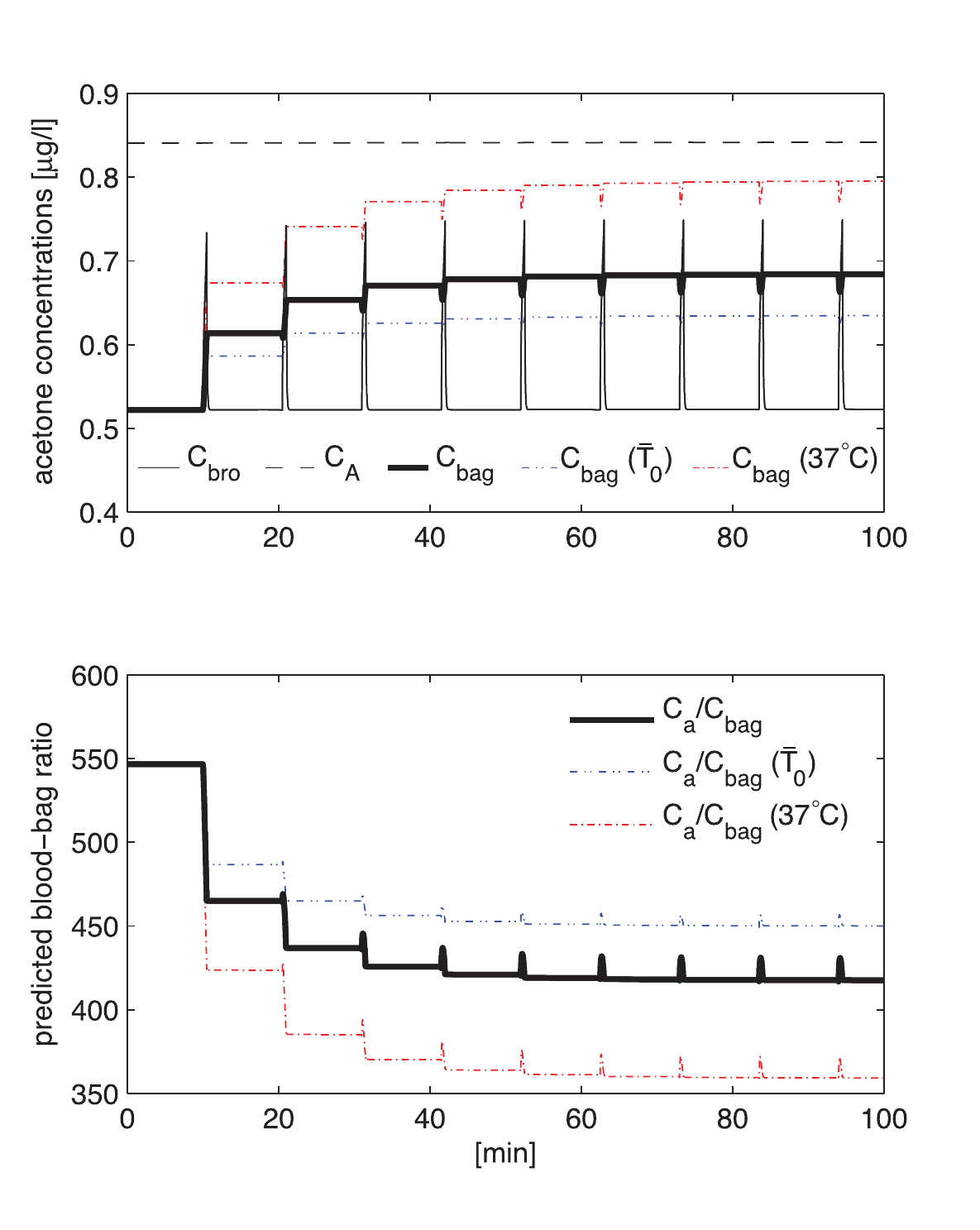}
\end{tabular}
\vspace*{-0.8cm}
\caption{Simulation of a sequential rebreathing protocol according to~\cite{ohara2008} with intermediate pauses of 10~min characterized by free tidal breathing. The underlying model parameters correspond to the individual fit in~\fref{fig:rebreathing_single}. Dash-dotted lines represent upper and lower bounds for the bag concentration as well as for the observable blood-bag ratio with respect to changes in airway temperature $\bar{T}$. These bounds were obtained by assuming that the airway temperature either instantaneously rises to body core temperature during the individual rebreathing periods or remains constant (i.e., at its initial level $\bar{T}_0$) throughout the entire experiment.}\label{fig:rebreathing_mult}
\end{figure}

From the data in the last two panels of~\fref{fig:rebreathing_single} it can be inferred that all physiological input variables will have returned to pre-rebreathing values within the 10~min breaks. For simulation purposes, it will hence be assumed that their behavior during repeated rebreathing segments is identical to the profiles within the first 0.5~min of single cycle rebreathing. Values for the initial compartment concentrations as well as for the additional parameters are adopted from the previous subsection. These premises allow for \emph{simulating} the evolution of breath acetone within a repeated rebreathing regime as displayed in~\fref{fig:rebreathing_mult}. Here, the initial bag concentration at the onset of each individual rebreathing cycle is determined by the final bag concentration after the preceding rebreathing cycle, i.e., no fresh room air enters the bag.

The bag concentration profile in~\fref{fig:rebreathing_mult} qualitatively resembles the data presented by~\cite{ohara2008}. However, what emerges from this modeling-based analysis is that in spite of steadily increasing bag concentrations (finally reaching a plateau level), the latter might not necessarily approach the underlying alveolar concentration as in the case of single cycle rebreathing. The major reason for this is a lack of complete thermal and diffusional equilibration between the airways and the alveolar region within the individual rebreathing segments. One potential way to circumvent this issue would be to reduce the desaturation and cooling of the airway tissues between consecutive rebreathing segments by keeping the intermediate time interval of free tidal breathing as short as possible (while simultaneously maintaining a regime allowing for comfortable breathing).

The second panel in~\fref{fig:rebreathing_mult} displays the evolution of the predicted blood-bag concentration ratios during the course of experimentation. Note that the in vitro blood:gas partition coefficient $\hen=340$ is never attained. This observation can offer some explanation for the discrepancies that continue to exist with regard to theoretical and experimentally measured ratios between blood and (rebreathed) breath levels~\cite{ohara2009}. Furthermore, the final plateau value and the observable BBR of acetone will vary with temperature (cf.~~\eref{eq:BBR}), which is consistent with similar observations made in the case of breath ethanol measurements~\cite{ohlsson1990}.

\section{Conclusions}\label{sect:dis}

Here we have successfully applied a previously published compartment model for the exhalation kinetics of highly soluble, blood-borne VOCs to the experimental framework of isothermal rebreathing. The proposed model has proven sufficiently flexible for capturing the corresponding end-tidal breath dynamics of acetone, which can be viewed as a prototypical test compound within this context. Moreover, it allows for a quantitative assessment of the discrepancies between in vitro and measured blood-breath ratios of such VOCs during the cyclical steady state observed in normal tidal breathing and can serve as a relatively simple mechanistic basis for describing some fundamental physiological processes associated with the observable VOC behavior in response to the above-mentioned experimental regime. \\

Several practical implications emerge from this modeling-based analysis. Firstly, it is demonstrated that the classical Farhi setting will fail to reproduce the experimentally measured acetone exhalation data during isothermal rebreathing if a constant endogenous production and metabolism rate is postulated. This is due to the fact that airway gas exchange, being a major determinant affecting the breath levels of highly soluble VOCs, is not taken into account within this formalism. Furthermore, multiplying end-tidal breath concentrations during free tidal breathing with the substance-specific blood:gas partition coefficient $\hen$ will generally underestimate the true arterial blood concentration for highly blood and water soluble VOCs.

Excessive hypoxia and hypercapnia are the main factors limiting the duration of the rebreathing maneuver and preventing a complete equilibration of VOC partial pressures in the alveoli and the conducting airways. From an operational point of view, our data indicate that even if isothermal rebreathing is continued until the individual breathing limit is reached, a quasi steady state according to~\eref{eq:rebreath} might not necessarily be attained. As can be deduced from~\fref{fig:rebreathing_single} in the case of acetone, end-exhaled breath (or bag) concentrations extracted after about 0.5~min of rebreathing (corresponding to the common protocol of providing around five consecutive rebreaths) are still likely to underestimate the underlying alveolar level $\calv$. Analogous conclusions can be drawn for sequential rebreathing protocols designed to allow for a recovery of the volunteer during the individual rebreathing segments.  

A reliable extraction of meaningful breath levels for highly soluble VOCs by virtue of isothermal rebreathing hence appears to require more sophisticated setups incorporating the continuous removal of CO$_2$ and replacement of metabolically consumed oxygen. Provided that the influence of chemical CO$_2$-absorption on the measured breath and bag concentrations is negligible, such setups might for instance be adapted from closed chamber techniques~\cite{filser1992} or general closed-circuit anesthesia systems. 

Alternatively, alveolar concentrations might be extrapolated to some extent from \emph{partially} equilibrated rebreathing samples, using numerical parameter estimation schemes for reconstructing VOC exhalation kinetics according to a given model structure. While further validation and data gathering needs to be carried out before such estimates can become practically relevant, the mechanistic descriptions discussed in this paper are intended as a first step towards achieving this goal.

\section*{Acknowledgements}
The research leading to these results has received funding from the
European Community Seventh Framework Programme~(FP7/2007-13) under
grant agreement No.~217967. We appreciate funding from the Austrian
Federal Ministry for Transport, Innovation and Technology (BMVIT/BMWA,
Project~818803, KIRAS). Gerald Teschl and Julian King acknowledge support from the Austrian Science Fund (FWF) under Grant No.~Y330. 
We greatly appreciate the generous support of the government of Vorarlberg, Austria. Furthermore, we are indebted to the reviewers for their constructive criticism of the manuscript.
 
\appendix
\section{Model equations and nomenclature}\label{sect:modeleq}
This appendix serves to give a roughly self-contained outline of the model structure sketched in~\fref{fig:model_struct}~\cite{King2010a}.
The time evolution of VOC concentrations is captured by taking into account standard conservation of mass laws for the individual compartments.

Local diffusion equilibria are assumed to hold at the air-tissue, tissue-blood and air-blood interfaces, the ratio of the corresponding concentrations being described by the appropriate partition coefficients $\lambda$, e.g., $\hen$. Unlike for low blood soluble compounds, the amount of highly soluble gas dissolved in the local blood volume of perfused compartments cannot generally be neglected, as it might significantly increase the corresponding compartmental capacities. This is particularly true for the airspace compartments. We hence use the \emph{effective} compartment volumes $\vbro:=V_{\mathrm{bro}}+V_{\mathrm{muc}}\lmuca$, 
$\valv:=V_{\mathrm{A}}+V_{\mathrm{c'}}\hen$, $\vliv:=V_{\mathrm{liv}}+V_{\mathrm{liv,b}}\lliv$ as well as $\vtis:=V_{\mathrm{tis}}$ and neglect blood volumes only for the mucosal and tissue compartment. 

Values for the individual compartment volumes, temperature-dependent partition coefficients, and physiological variables such as cardiac output ($\qc$) and alveolar ventilation ($\qalv$) are given in~\tref{table:param}. Fractional blood flows to the bronchial tract ($\qbr$) and the liver ($\qliv$) as well as the endogenous production ($\pr$) and metabolic elimination rates ($\met$) are treated as constants in the context of isothermal rebreathing.

According to~\fref{fig:model_struct}, the mass balance equation for the bronchial compartment reads
\begin{equation}\label{eq:bro}
\frac{\di\cbr}{\di t}\vbro=\qalv(\cinh-\cbr)+D(\calv-\cbr)\\+\qbr\qc\big(\cart-z(\bar{T})\hen\cbr\big),
\end{equation}
where $\cinh$ denotes the inhaled (ambient/bag) gas concentration. Similarly, for the alveolar, liver and tissue compartment we find that
\begin{equation}\label{eq:alv}
\frac{\di\calv}{\di t}\valv=D(\cbr-\calv)+(1-\qbr)\qc\big(\cven -\hen\calv\big),
\end{equation} 
and
\begin{equation}\label{eq:liv}
\frac{\di\cliv}{\di t}\vliv=\pr-\met\lliv\cliv+\qliv(1-\qbr)\qc\big(\cart-\lliv\cliv\big),
\end{equation}
and
\begin{equation}\label{eq:tis}
\frac{\di\ctis}{\di t}\vtis=(1-\qliv)(1-\qbr)\qc\big(\cart-\ltis\ctis\big),
\end{equation}
respectively. Here,
\begin{equation}\label{eq:cven}\cven:=\qliv\lliv\cliv+(1-\qliv)\ltis\ctis\end{equation}
and
\begin{equation}\label{eq:cart}\cart:=(1-\qbr)\hen\calv+\qbr z(\bar{T})\hen\cbr\end{equation}
are the associated concentrations in mixed venous and arterial blood, respectively.


\setcounter{table}{0}
\begin{table}[H]
\centering 
\footnotesize
\begin{tabular}{|lcr|}\hline
 {\large\strut}Parameter&Symbol&Nominal value (units)\\ \hline \hline
{\large\strut}\textit{Compartment volumes} & & \\
{\large\strut}	Bronchioles &$V_{\mathrm{bro}}$ & 0.1 (l)$^a$\\
{\large\strut}	Mucosa &$V_{\mathrm{muc}}$ & 0.005 (l)$^a$\\
{\large\strut}	Alveoli &$V_{\mathrm{A}}$ & 4.1 (l)$^a$\\
{\large\strut}	End-capillary &$V_{\mathrm{c'}}$ & 0.15 (l)$^b$\\
{\large\strut}	Liver &$V_{\mathrm{liv}}$ & 0.0285\,LBV (l)$^a$\\
{\large\strut}	Blood liver &$V_{\mathrm{liv,b}}$ & 1.1 (l)$^c$\\
{\large\strut}	Tissue &$V_{\mathrm{tis}}$ & 0.7036\,LBV (l)$^a$\\
{\large\strut}	Rebreathing bag &$\vbag$ & 3 (l)$^a$\\
{\large\strut}\textit{Respiratory parameters} & & \\
{\large\strut}  Breathing frequency &$f$ & 12.5 (tides/min)$^d$\\
{\large\strut}  Tidal volume &$V_T$ & 0.593 (l)$^d$\\
{\large\strut}  Alveolar ventilation &$\qalv$ & 6.2 (l/min)$^e$\\
{\large\strut}\textit{Blood flows} & & \\
{\large\strut}  Cardiac output &$\qc$ & 6 (l/min)$^f$\\
{\large\strut}	Fractional flow bronchioles &$\qbr$ & 0.01$^g$\\
{\large\strut}	Fractional flow liver&$\qliv$ & 0.32$^a$\\
{\large\strut}\textit{Partition coefficients} (37$^\circ$C) & & \\
{\large\strut}	Blood:air &$\hen$ & 340$^h$\\
{\large\strut}  Mucosa:air &$\lmuca$ & 392$^{i,j}$\\
{\large\strut}  Blood:liver &$\lliv$ & 1.73$^j$\\
{\large\strut}  Blood:tissue &$\ltis$ & 1.38$^h$\\
{\large\strut}\textit{Metabolic and diffusion constants} & & \\
{\large\strut}	Linear metabolic rate &$\met$ & 0.0074 (l/kg$^{\textrm{0.75}}$/min)$^l$\\
{\large\strut}  Endogenous production &$\pr$ & 0.19 (mg/min)$^l$\\
{\large\strut}  Stratified conductance &$D$ & 0 (l/min)$^l$\\
{\large\strut}  Constant~\eref{eq:D} &$\dke$ & 14.9 (min$^{-1}$)$^l$\\
{\large\strut}  Constant~\eref{eq:D} &$\dkz$& 0.76$^l$\\
\hline
\end{tabular}
\caption{Basic model parameters and nominal values during rest. LBV denotes the lean body volume in liters calculated according to $\textrm{LBV}=-16.24+0.22\,\mathrm{bh}+0.42\,\mathrm{bw}$, with body height ($\mathrm{bh}$) and weight ($\mathrm{bw}$) given in cm and kg, respectively~\cite{moerk2006}; $^a$\cite{moerk2006}; $^b$\cite{pulmcirc}; $^c$\cite{ottesen2004}; $^d$\cite{duffin2000}; $^e$cf.~\eref{eq:qalv}; $^f$\cite{mohrman2006}; $^g$\cite{lumbbook}; $^h$\cite{anderson2006};
 $^i$\cite{staudinger2001}; $^j$\cite{kumagai1995}; $^l$\cite{King2010a}.}\label{table:param}
\end{table}

\newpage
\section*{References}
\bibliographystyle{unsrt}

\bibliography{AllCit}

\end{document}